\documentclass[preprint,prl,onecolumn]{revtex4}

\usepackage{latexsym}
\usepackage{amssymb}
\usepackage{amsmath}
\usepackage{graphicx}

\begin{document}


\title{Control and detection of
singlet-triplet mixing in a random nuclear field }

\author{F. H. L. Koppens}
\affiliation{Kavli Institute of NanoScience Delft, P.O. Box 5046,
2600 GA Delft, The Netherlands}

\author{J. A. Folk}
\affiliation{Kavli Institute of NanoScience Delft, P.O. Box 5046,
2600 GA Delft, The Netherlands}

\author{J. M. Elzerman}
\affiliation{Kavli Institute of NanoScience Delft, P.O. Box 5046,
2600 GA Delft, The Netherlands}

\author{R. Hanson}
\affiliation{Kavli Institute of NanoScience Delft, P.O. Box 5046,
2600 GA Delft, The Netherlands}

\author{L. H. Willems van Beveren}
\affiliation{Kavli Institute of NanoScience Delft, P.O. Box 5046,
2600 GA Delft, The Netherlands}

\author{I. T. Vink}
\affiliation{Kavli Institute of NanoScience Delft, P.O. Box 5046,
2600 GA Delft, The Netherlands}

\author{H. P. Tranitz}
\affiliation{Institut f\"ur Angewandte und Experimentelle Physik,
Universit\"at Regensburg, Regensburg, Germany}

\author{W. Wegscheider}
\affiliation{Institut f\"ur Angewandte und Experimentelle Physik,
Universit\"at Regensburg, Regensburg, Germany}

\author{L. P. Kouwenhoven}
\affiliation{Kavli Institute of NanoScience Delft, P.O. Box 5046,
2600 GA Delft, The Netherlands}

\author{L. M. K. Vandersypen}
\affiliation{Kavli Institute of NanoScience Delft, P.O. Box 5046,
2600 GA Delft, The Netherlands}



\begin{abstract}
  We observe mixing between two-electron singlet and triplet states
in a double quantum dot, caused by interactions with nuclear spins
in the host semiconductor. This mixing is suppressed by applying a
small magnetic field, or by increasing the interdot tunnel
coupling and thereby the singlet-triplet splitting. Electron
transport involving transitions between triplets and singlets in
turn polarizes the nuclei, resulting in striking bistabilities. We
extract from the fluctuating nuclear field a limitation on the
time-averaged spin coherence time $T_2^*$ of 25 ns. Control of the
electron-nuclear interaction will therefore be crucial for the
coherent manipulation of individual electron spins.
\end{abstract}

\maketitle


A single electron confined in a GaAs quantum dot is often referred
to as artificial hydrogen. One important difference between
natural and artificial hydrogen, however, is that in the first,
the hyperfine interaction couples the electron to a single
nucleus, while in artificial hydrogen the electron is coupled to
about a million Ga and As nuclei. This creates a subtle interplay
between electron spin eigenstates affected by the ensemble of
nuclear spins (the Overhauser shift), nuclear spin states affected
by time-averaged electron polarization (the Knight shift), and the
flip-flop mechanism that trades electron and nuclear spins
\cite{opticalorientation,Gammon}.

The electron-nuclear interaction has important consequences for
quantum information processing with confined electron spins
\cite{Loss-Divincenzo}. Any randomness in the Overhauser shift
introduces errors in a qubit state, if no correcting measures are
taken \cite{Merkulov,Braun,Bracker}. Even worse, multiple qubit
states, like the entangled states of two coupled electron spins,
are redefined by different Overhauser fields. Characterization and
control of this mechanism will be critical both for identifying
the problems as well as finding potential solutions.

Here, the implications of the hyperfine interaction on entangled
spin states are studied in two coupled quantum dots --- an
artificial hydrogen molecule --- where the molecular states can be
controlled electrically.  A random polarization of nuclear spins
creates an inhomogeneous effective field that couples molecular
singlet and triplet states, and leads to new eigenstates that are
admixtures of these two. We use transport measurements to
determine the degree of mixing over a wide range of tunnel
coupling, and observe a subtle dependence of this mixing on
magnetic field. We find that we can controllably suppress the
mixing by increasing the singlet-triplet splitting. This ability
is crucial for reliable two-qubit operations such as the SWAP gate
\cite{Loss-Divincenzo}.

Furthermore, we find that electron transport itself acts back on
the nuclear spins through the hyperfine interaction, and
time-domain measurements reveal complex, often bistable, behavior
of the nuclear polarization.  Understanding the current-induced
nuclear polarization is an important step towards electrical
control of the nuclear spins.  Such control will be critical for
electrical generation and detection of entangled nuclear spin
states \cite{Eto}, or for transfer of quantum information between
electron and nuclear spin systems \cite{Kane, Lukin}. Even more
appealing will be reducing the nuclear field fluctuations in order
to achieve longer electron spin coherence times
\cite{Schlieman,Khaetskii,Coish}.

The coupled electron-nuclear system is studied using electrical
transport measurements through two dots in series \cite{Wiel}, in
a regime where the Pauli exclusion principle blocks current flow
\cite{Ono,Johnson}. The dots are defined using electrostatic gates
on a GaAs/AlGaAs heterostructure (see Fig.\,1E and \cite{stats}).
The gate voltages are tuned such that one electron always resides
in the right dot, and current flows if a second electron tunnels
from the left reservoir, through the left and right dots, to the
right reservoir (see Fig.\,1D).  This cycle can be described using
the occupations $(m,n)$ of the left and right dots:
$(0,1)\rightarrow(1,1)\rightarrow(0,2)\rightarrow(0,1)$.  When an
electron enters from the left dot, the two-electron system forms
either a molecular singlet, S(1,1), or a molecular triplet,
T(1,1). From S(1,1), the electron in the left dot can move to the
right dot to form S(0,2). From T(1,1), however, the transition to
(0,2) is forbidden by spin conservation (T(0,2) is much higher in
energy than S(0,2)). Thus, as soon as T(1,1) is occupied, further
current flow is blocked (Pauli blockade).

A characteristic measurement of this blockade is shown in
Fig.\,1A. The suppression of current ($<80$ fA) in the region
defined by dashed lines is a signature of Pauli blockade
\cite{Ono,Johnson} (see also Fig.\,S1 and \cite{SOM}). Fig.\,1B
shows a similar measurement, but with a much weaker interdot
tunnel coupling $t$. Strikingly, a large leakage current now
appears in the Pauli blockaded region, even though the barrier
between the two dots is more opaque. Furthermore, this leakage
current is substantially reduced by an external magnetic field of
only 100 mT (Fig.\,1C). Such a strong field dependence is
remarkable because the in-plane magnetic field, $B_{ext}$, couples
primarily to spin and the Zeeman energies involved are very small
($E_Z\sim 2.5\,\mu$eV at $B_{ext} =$100 mT compared to the thermal
energy, $\sim15\,\mu$eV at 150 mK, for example).

Leakage in the Pauli blockade regime occurs when singlet and
triplet states are coupled. The T(1,1) state that would block
current can then transition to the S(1,1) state and the blockade
is lifted (Fig.\,1D).  As we will show, coupling of singlets and
triplets in Figs.\,1B,C originates from the hyperfine interaction
between the electron spins and the Ga and As nuclear spins (other
leakage mechanisms can be ruled out, see \cite{SOM}).

The hyperfine interaction between an electron with spin
$\overrightarrow{S}$ and a nucleus with spin $\overrightarrow{I}$
has the form $A\overrightarrow{I}\cdot\overrightarrow{S}$, where
$A$ characterizes the coupling strength. An electron coupled to an
ensemble of $n$ nuclear spins experiences an effective magnetic
field $\overrightarrow{B_N} \sim \frac{1}{g \mu_B} \sum_i^n A_i
\overrightarrow{I_i}$, with $g$ the electron $g$-factor and
$\mu_B$ the Bohr magneton \cite{opticalorientation}. For fully
polarized nuclear spins in GaAs, $B_N\sim 5$ T \cite{Paget77}. For
unpolarized nuclear spins, statistical fluctuations give rise to
an effective field pointing in a random direction with an average
magnitude of $5$ T$/\sqrt{n}$ \cite{Merkulov,GammonPRL,Braun}.
Quantum dots like those measured here contain $n\sim 10^6$ nuclei,
so $\| \overrightarrow{B_N} \| \sim5$ mT.

Nuclei in two different dots give rise to effective nuclear
fields, $\overrightarrow{B_{N1}}$ and $\overrightarrow{B_{N2}}$,
that are uncorrelated. Although the difference in field
$\overrightarrow{\Delta
B_{N}}=\overrightarrow{B_{N1}}-\overrightarrow{B_{N2}}$ is small,
corresponding to an energy $E_N\equiv g \mu_B \| \Delta
\overrightarrow{B_N} \| \sim 0.1\mu$eV, it nevertheless plays a
critical role in Pauli blockade. The (1,1) triplet state that
blocks current flow consists of one electron on each of the two
dots. When these two electrons are subject to different fields,
the triplet is mixed with the singlet and Pauli blockade is
lifted. For instance, an inhomogeneous field along  $\hat{z}$
causes the triplet $|T_0\rangle =
\frac{1}{\sqrt{2}}(|\uparrow\downarrow\rangle +
|\downarrow\uparrow\rangle)$ to evolve into the singlet
$\frac{1}{\sqrt{2}}(|\uparrow\downarrow\rangle -
|\downarrow\uparrow\rangle)$. Similarly, the other two triplet
states, $|T_+\rangle = |\uparrow\uparrow\rangle$ and $|T_-\rangle
= |\downarrow\downarrow\rangle$, evolve into the singlet due to
$\hat{x}$ and $\hat{y}$ components of $\overrightarrow{\Delta
B_{N}}$.

The degree of mixing by the inhomogeneous field depends on the
singlet-triplet energy splitting, $E_{ST}$. Singlet and triplet
states that are close together in energy ($E_{ST} \ll E_N$) are
strongly mixed, while states far apart in energy ($E_{ST} \gg
E_N$) experience only a slight perturbation due to the nuclei.

The singlet-triplet splitting depends on the interdot tunnel
coupling $t$ and on the detuning of left and right dot potentials
$\Delta_{LR}$.  $\Delta_{LR}$ and $t$ are controlled
experimentally using gate voltages (Fig.\,1E).  $V_t$ controls the
interdot tunnel coupling.  $V_L$ and $V_R$ set the detuning, and
thereby determine whether transport is inelastic (detuned levels),
resonant (aligned levels), or blocked by Coulomb blockade
(Fig.\,1F). The coupling of the dots to the leads is held constant
using $V_{lead}$.

The effect of the two tunable parameters $t$ and $\Delta_{LR}$ on
the singlet and triplet energies is illustrated in Figs.\,2A and
2B.  For weak tunnel coupling ($t\sim 0$), and in the absence of a
hyperfine interaction ($E_N\sim 0$), the (1,1) singlet and (1,1)
triplet states are nearly degenerate (Fig.\,2A). A finite interdot
tunnel coupling $t$ leads to an anticrossing of S$(1,1)$ and
S$(0,2)$. The level repulsion results in an increased
singlet-triplet splitting that is strongly dependent on detuning
(Fig.\,2B).  At the resonant condition ($\Delta_{LR}=0$, aligned
levels), the two new singlet eigenstates are equidistant from the
triplet state, both with $E_{ST}=\sqrt{2}t$.  For finite detuning
(but smaller than the single dot S-T splitting), one singlet state
comes closer to the triplet state ($E_{ST}\sim t^2/\Delta_{LR}$),
while the other moves away. Both in Figs.\,2A and 2B, singlet and
triplet states are pure eigenstates (not mixed) and therefore
Pauli blockade would be complete.

The additional effect of the inhomogeneous nuclear field is shown
in Figs.\,2C and 2D. For small $t$ ($\sqrt{2}t, t^2/\Delta_{LR} <
E_N$), the (1,1) singlet and (1,1) triplet are close together in
energy and hence mix strongly (purple lines) over the entire range
of detuning. For $t$ such that $t^2/\Delta_{LR} < E_N <
\sqrt{2}t$, triplet and singlet states mix strongly only for
finite detuning. This is because $E_{ST}$ is larger than $E_N$ for
aligned levels but smaller than $E_N$ at finite detuning. For
still larger $t$ ($\sqrt{2}t, t^2/\Delta_{LR} > E_N$, not shown in
Fig.\,2), mixing is weak over the entire range of detuning. In the
cases where mixing between S and T is strong, as in Figs.\,2C and
D (for large detuning), Pauli blockade is lifted and a leakage
current results.

The competition between $E_{ST}$ and $E_N$ can be seen
experimentally by comparing 1D traces of leakage current as a
function of detuning over a wide range of $t$ (Fig.\,3A). Resonant
current appears as a peak at $\Delta_{LR}=0$; inelastic leakage as
the shoulder at $\Delta_{LR}>0$ \cite{Fujisawa}. When the interdot
tunnel coupling is small, both resonant and inelastic transport
are allowed due to singlet-triplet mixing, and both rise as the
middle barrier becomes more transparent. As the tunnel coupling is
raised further, the point is reached where $E_{ST}$ becomes larger
than the nuclear field and Pauli blockade suppresses the current
(see also Fig.\,1A). The maximum resonant current occurs at a
smaller value of $t$ compared to the maximum inelastic current
(see inset in Fig.\,3A). This is consistent with $E_{ST}$ being
much smaller for finite detuning than for aligned levels,
$t^2/\Delta_{LR} \ll \sqrt{2}t$ (Figs.\,2B, D).

The experimental knob provided by electrostatic gates is very
coarse on the energy scales relevant to the hyperfine interaction.
However, the external magnetic field can easily be controlled with
a precision of $0.1$ mT, corresponding to a Zeeman splitting (2
neV) that is fifty times smaller than $E_N$. For this reason,
monitoring the field dependence allows a more detailed examination
of the competing energy scales $E_{ST}$, $E_Z$ and $E_N$.

The competition between $E_Z$ and $E_N$ is clear for small
interdot tunnel coupling (Fig.\,3B). Leakage current is suppressed
monotonically with magnetic field, on a scale of $\sim5$ mT and
$\sim10$ mT for inelastic and resonant transport respectively. The
qualitative features of this field dependence can be understood
from the insets in Fig.\,2C. At zero field all states are mixed
strongly by the inhomogeneous nuclear field, but when $E_Z$
exceeds $E_N$, the mixing between the singlet and two of the
triplet states ($|T_+\rangle$ and $|T_-\rangle$) is suppressed. An
electron loaded into either of these blocks further current flow
and leakage disappears.

The magnitude of the fluctuating Overhauser field can be extracted
from the inelastic peak shape in the limit of small $t$ (as in the
inset of Fig.\,3B). A fit of the data with a model that describes
the transport cycle using the density matrix approach
\cite{Nazarovpriv} is presented in \cite{SOM}. From this fit, we
find a magnitude of the inhomogeneous field $\sqrt{\langle \Delta
{B_N}^2\rangle}=1.73 \pm 0.02 $ mT ($E_N= 0.04\,\mu$eV), largely
independent of $\Delta_{LR}$ over the parameter range studied
\cite{leadbarriers}. The value for the effective nuclear field
fluctuations in a single dot is obtained from the relation
${\langle {B_N}^2\rangle}=\frac{1} 2 \langle \Delta
{B_N}^2\rangle$, giving $\sqrt{\langle {B_N}^2\rangle}=1.22$ mT.
This is consistent with the strength of the hyperfine interaction
in GaAs and the number of nuclei that are expected in each dot
\cite{Merkulov, JohnsonRelax}.

The three-way interplay between $E_{ST}$, $E_Z$ and $E_N$ is most
clearly visible in the resonant current.  At an intermediate value
of tunnel coupling, $t \gtrsim E_N$ (Fig.\,3C), the resonant peak
is split in magnetic field, with maxima at $\pm 10$ mT (see
inset). This behavior can be understood from the lower inset in
Fig.\,2D. At $B_{ext} = 0$, the current is somewhat suppressed
compared to Fig.\,3B, because now $E_{ST} > E_N$. Increasing
$B_{ext}$ enhances the mixing as the $|T_+\rangle$ and
$|T_-\rangle$ states approach the singlet states. The maximum
leakage occurs when the states cross, at $E_{ST} (= \sqrt{2}t) =
E_Z$. Here, $E_Z=0.25\pm 0.03\,\mu$eV at the current maximum, from
which we can extract $t=0.18\pm 0.02 \mu$eV for this setting of
$V_t$. At still larger $B_{ext}$, $|T_+\rangle$ and $|T_-\rangle$
move away from the singlet states again, and the leakage current
is suppressed.

The system enters into a new regime for still higher tunnel
coupling (Figs.\,3D and 4), where it becomes clear that the
electron-nuclear system is dynamic.   The zero field resonant
leakage is further suppressed and above $10$ mT prominent current
spikes appear (left inset).  The spikes are more dramatically
visible in a 3D surface plot of leakage over a broader range of
field (Fig.\,4A). Even for fixed experimental parameters, the
current fluctuates in time as shown in Fig.\,4B.

We find that time dependent behavior is a consistent feature of
resonant transport for \\ $(E_{ST},E_Z) \gg E_{N}$. For some
settings the time dependence is fast (see, e.g., the fluctuations
of Figs.\,4A and 4B), but for others the leakage changes much more
slowly.  An example of the slower time dependence is shown in
Fig.\,4C.  Starting from an equilibrium situtation (bias voltage
switched off for five minutes), the current is initially very
small after the bias is turned on. It builds up and then saturates
after a time that ranges from less than a second to several
minutes.  This timescale depends on $\Delta_{LR}$, $t$, and
$B_{ext}$. When no voltage bias is applied, the system returns to
equilibrium after $\sim 80$ s at $200$ mT. Similar long timescales
of the nuclear spin-lattice relaxation times have been reported
before in GaAs systems \cite{Dobers} and quantum dots
\cite{Blick}. We thus associate these effects with current-induced
dynamic nuclear polarization and relaxation.

Evidence that the fast fluctuations too are related to current
induced nuclear polarization (and cannot be explained by
fluctuating background charges alone), is found in their
dependence on sweep direction and sweep rate \cite{Dobers,temp}.
When the magnetic field is swept while maintaining fixed
$\Delta_{LR}$, the current shows fluctuations at low field but
suddenly becomes stable at high field (Fig.\,4D).  The crossover
from unstable to stable behavior occurs at a field that is
hysteretic in sweep direction (Fig.\,4D), and this hysteresis
becomes more pronounced at higher sweep rates (faster than $\sim
1$ mT/s). The connection between the fluctuations and nuclear
polarization is also evident from time traces, where instability
develops only after the nuclear polarization is allowed to build
for some time (Fig.\,S3).

Unlike the regular oscillations that have been observed in other
GaAs structures (see e.g. \cite{OnoPRL,opticalorientation}), the
fluctuations in our measurements are random in time, and in many
cases suggest bistability with leakage current moving between two
stable values. We discuss the origin of such fast bistable
fluctuations in \cite{SOM}.

The ensemble of random nuclear spins that gives rise to the mixing
of two-electron states as observed in this experiment also gives
rise to an uncertainty of $g \mu_B \sqrt{\langle {B_N}^2 \rangle}
=0.03\,\mu$eV in the Zeeman energy of one electron. When averaged
over a time longer than the correlation time of the nuclear spin
bath ($\sim 100\,\mu$s) \cite{Sousa}, this implies an upper limit
on the dephasing time of $T_2^* = \hbar / {g \mu_B \sqrt{ \frac{2}
3 \langle {B_N}^2 \rangle}} =25$ ns (following the definition of
\cite{Merkulov}), comparable to the $T_2^*$ found in recent
optical spectroscopy measurements \cite{T2comparison}. This value
is four orders of magnitude shorter than the theoretical
prediction for the electron spin $T_2$ in the absence of nuclei,
which is limited only by spin-orbit interactions
\cite{Golovach,Elzerman2,Kroutvar}.

One way to eliminate the uncertainty in Zeeman energy that leads
to effective dephasing is to maintain a well-defined nuclear spin
polarization \cite{Coish}. Many of the regimes explored in this
paper show leakage current that is stable when current-induced
polarization is allowed to settle for some time. These may in fact
be examples of specific nuclear polarizations that are maintained
electrically.

\vspace{10mm} \noindent
{\bf Supporting online text and figures}\\
(www.sciencemag.org/cgi/content/full/1113719/DC1)\\

\clearpage
\bibliography{KoppensBib}

\begin{thebibliography}{33}
\bibitem[{sta()}]{stats}
\bibinfo{note}{The electrostatic gates are on the surface of a GaAs/AlGaAs
  heterostructure. The 2DEG is 90 nm below the surface, with density
  $1.33\times 10^{11}\,cm^{-2}$ and mobility $9.71\times 10^{5}\,cm^2/Vs$.
  Measurements are performed in a dilution refrigerator at 150mK, with a
  magnetic field in the plane of the heterostructure.}
\bibitem[{SOM()}]{SOM}
\bibinfo{note}{Supporting text is available on Science Online.}
\bibitem[{lea()}]{leadbarriers}
\bibinfo{note}{The difference between the field dependence of the resonant and
  inelastic current can be explained by the coupling with the leads. The S(0,2)
  state is lifetime broadened due to coupling with the right lead ($\sim
  0.3\,\mu$eV) giving a weaker field dependence for the resonant current. The
  field dependence of the inelastic leakage is not affected by the lead
  coupling, because under the high bias conditions of this experiment, there
  are no available states in the left lead that could broaden S(1,1).}
\bibitem[{tem()}]{temp}
\bibinfo{note}{The leakage current is sensitive to magnetic fields of only a
  few mT, corresponding to $0.1\%$ nuclear polarization. Given that the dot has
  $\sim 10^6$ nuclei, changes in the nuclear polarization of $0.1\%$ can be
  caused by 1000 electron-nuclear flip-flop processes. For typical currents
  ($\sim 100$ fA) 1000 electrons move through the dot in one millisecond, so in
  principle current fluctuations as fast as $1$ kHz are possible.}
\bibitem[{T2c()}]{T2comparison}
\bibinfo{note}{In Ref. \cite{Bracker}, a $T_2^*$ of 16 ns was found in a GaAs
  quantum dot, slightly shorter than in our experiment, presumably due to a
  smaller dot size. A considerably shorter timescale, 500 ps, was measured in
  InAs dots \cite{Braun}, due to a notable difference in dot size, nuclear spin
  $I$, and the stronger hyperfine coupling constant in InAs.}
\end{thebibliography}
\bibliographystyle{apsrev}

\cite{ack}


\pagestyle{empty}

\begin{figure}[htbf]
\begin{center}
\includegraphics[width=5in]{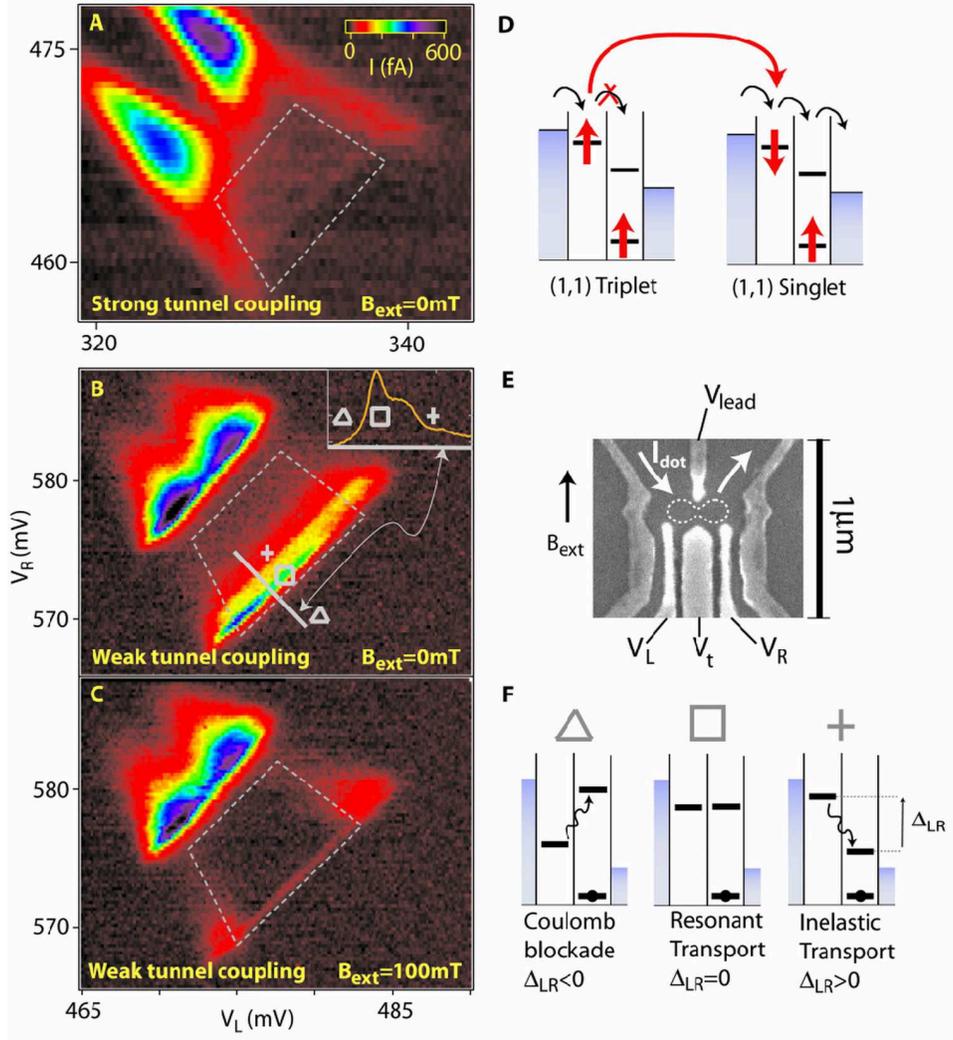}
\end{center}

\caption{\addtolength{\baselineskip}{-8pt} Pauli
blockade and leakage current. ({\bf A}) Color-scale plot of the
current through two coupled dots as a function of the left and
right dot potentials (voltage bias 800 $\mu$eV, $V_t = -108$ mV).
The experimental signature of Pauli blockade is low current ($<80$
fA) in the area denoted by dotted grey lines. ({\bf B}) Analogous
data for smaller interdot tunnel coupling ($V_t=-181$ mV), with
the same color scale as in (A). A dramatic increase of leakage
current is seen in the lower part of the Pauli blockaded area
(green/yellow band). Inset: 1D trace along the solid grey line,
with Coulomb blockaded, resonant and inelastic transport regimes
marked, see also (F). ({\bf C}) Analogous data for the same tunnel
coupling as in (B), but for $B_{ext} = 100$ mT. The leakage
current from (B) is strongly suppressed. ({\bf D}) Two level
diagrams that illustrate Pauli blockade in coupled quantum dots
(see text). When the (1,1) triplet is changed into the (1,1)
singlet (red arrow), Pauli blockade is lifted. ({\bf E}) SEM
micrograph showing the device geometry. White arrows indicate
current flow through the two coupled dots (dotted line). ({\bf F})
Level diagrams illustrating three transport regimes. $\triangle$:
Coulomb blockade; transport would require absorption of energy.
$\Box$: Resonant transport; the dot levels are aligned. $+$:
Inelastic transport; energy must be transferred to the
environment, for instance by emitting a phonon. These symbols are
used to denote inelastic, resonant, and Coulomb blockade regimes
in Figs.\,1, 2, and 3.}
\end{figure}

\begin{figure}[h]
\begin{center}
\includegraphics[width=5in]{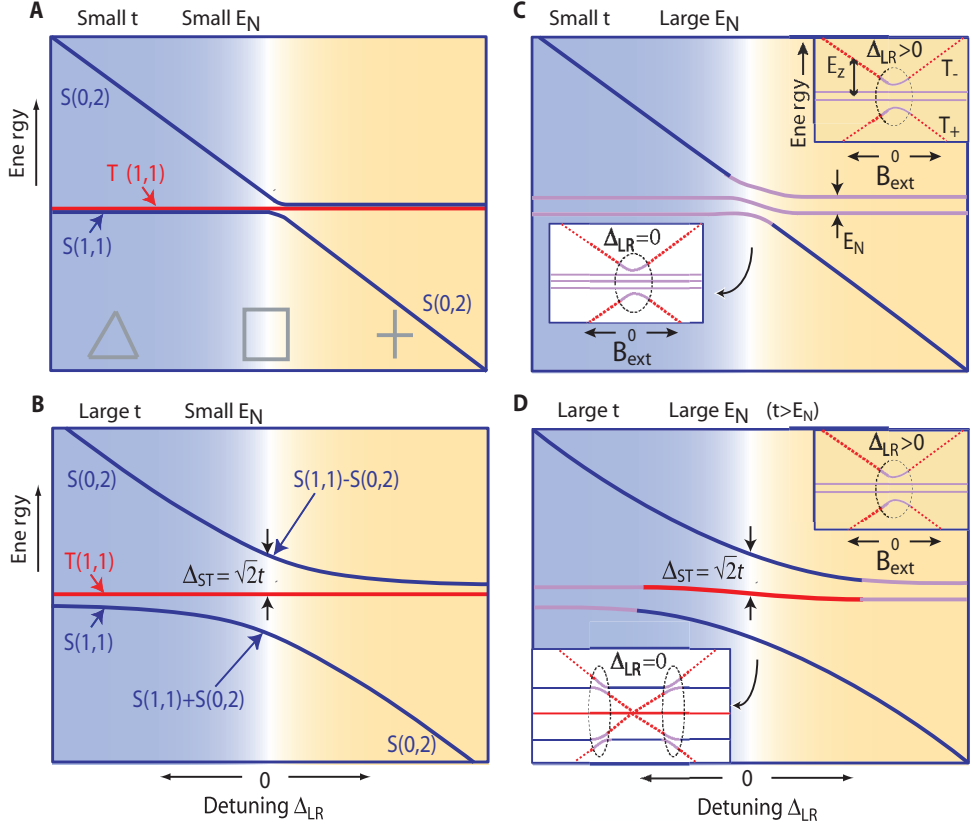}
\end{center}
\caption{\addtolength{\baselineskip}{-8pt}
Two-electron level diagrams showing energy as a function of
detuning $\Delta_{LR}$. Detuning is defined so that the energy of
T(1,1) remains constant as $\Delta_{LR}$ varies (see also
Fig.\,S1B and the supporting online text). The panels on the left
illustrate the effect of $t$; the panels on the right include the
additional effect of an inhomogeneous magnetic field. Pure singlet
and triplet states are drawn in blue and red respectively, strong
admixtures in purple. The blue, white and yellow background
correspond to the Coulomb blockade, resonant and inelastic
transport regime respectively. ({\bf A}) For small tunnel
coupling, T(1,1) and S(1,1) are nearly degenerate. ({\bf B}) For
finite $t$, level repulsion between the singlet states results in
a larger singlet-triplet splitting compared to (A), which depends
on detuning. The tunnel coupling does not mix singlet and triplet
states. For large $\Delta_{LR}$ (but smaller than the single dot
S-T splitting) $E_{ST}\sim t^2/\Delta_{LR}$. ({\bf C} and {\bf D})
An inhomogeneous field mixes triplet and singlet states that are
close in energy (purple lines). For clarity only one triplet state
is shown in the main panels. ({\bf C}) For small $t$, T(1,1) and
S(1,1) mix strongly over the full range of detuning. ({\bf D}) For
large $t$, T(1,1) mixes only strongly with the singlet for large
detuning. The insets to (C) and (D) show the effect of an external
magnetic field on the two-electron energy levels. All three
triplets are shown now; the triplets $|T_+\rangle$ and
$|T_-\rangle$ split off from $|T_0\rangle$ due to $B_{ext}$. The
leakage current is highest in the regions indicated by black
dotted ellipses (see text).}
\end{figure}

\begin{figure}[h]
\begin{center}
\includegraphics[width=2.0in]{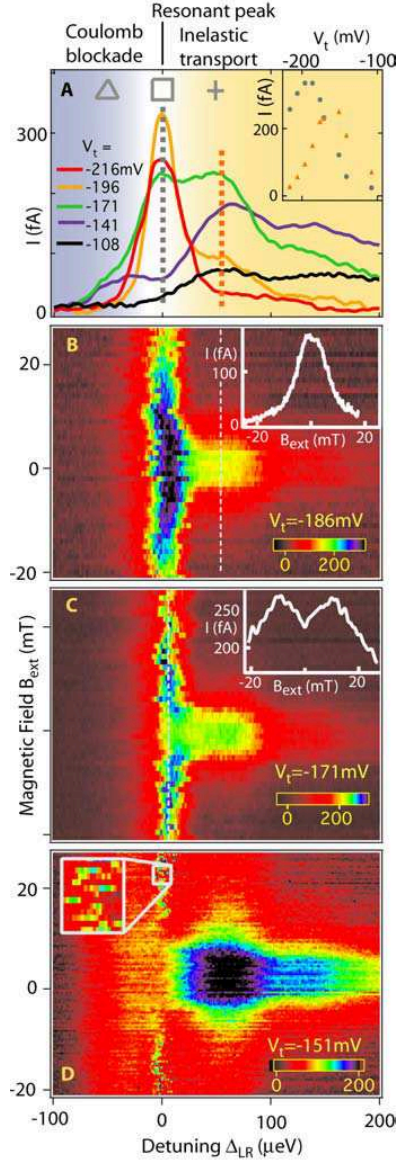}

\caption{\addtolength{\baselineskip}{-8pt} The
measured leakage current results from a competition between $E_N$,
$E_{ST}$ and $E_Z$. ({\bf A}) 1D traces of the leakage current as
a function of detuning at $B_{ext}=0$, for a wide range of tunnel
couplings (analogous to the inset of Fig.\,1B). Coulomb blockade,
resonant transport and inelastic transport are indicated by
colored backgrounds as in Fig.\,2. Inset: leakage current along
the dotted grey and orange lines is shown as a function of $V_t$.
Resonant and inelastic leakage (grey and orange markers) reach a
maximum at different tunnel couplings ($V_t=-190$ mV and $-150$ mV
respectively). ({\bf B}) For small tunnel coupling ($< E_N$), both
the resonant and inelastic leakage currents drop monotonically
with $B_{ext}$. Inset: magnetic field dependence of the inelastic
current along the dotted line ($\Delta_{LR}=40\,\mu$eV).  ({\bf
C}) For larger $t$ ($>E_N$), the resonant leakage current is
maximum at $B_{ext} \pm 10$ mT. Inset: field dependence of the
resonant peak height (dotted line). ({\bf D}) For still larger
$t$, the resonant current is strongly reduced at low field, then
becomes unstable for higher field (see inset).}
\end{center}
\end{figure}

\begin{figure}[h]
\begin{center}
\includegraphics[width=5in]{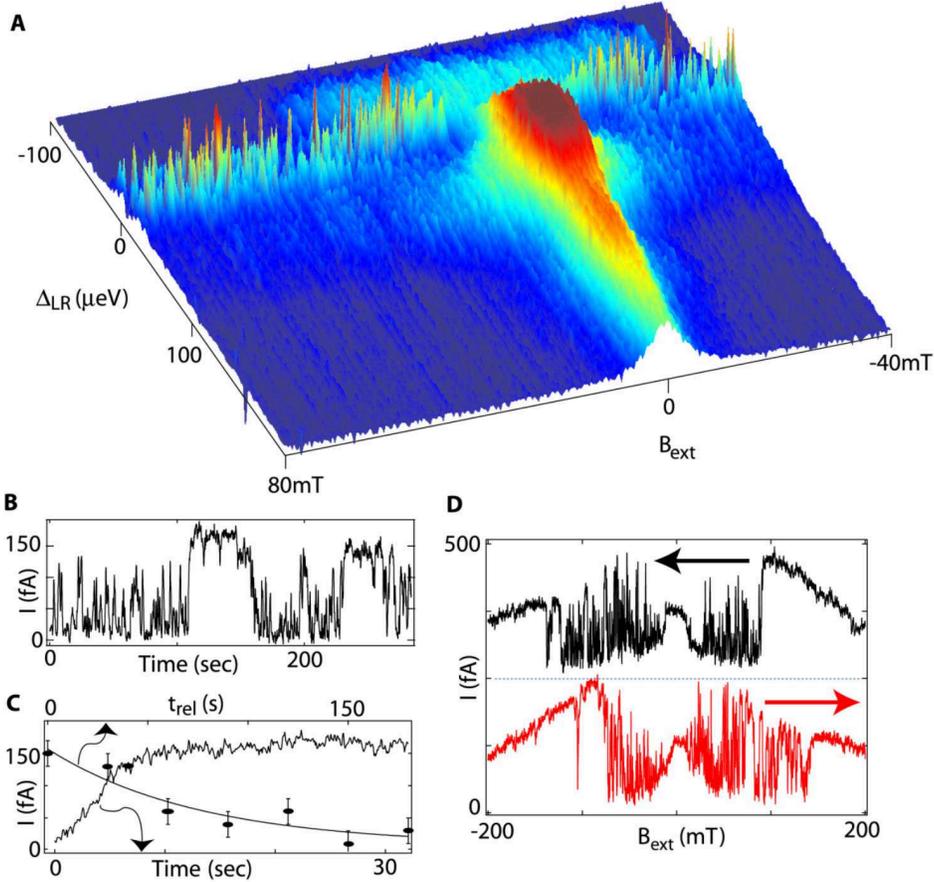}

\caption{ \addtolength{\baselineskip}{-8pt} Time dependence of the
leakage current reveals the dynamics of the electron-nuclear
system. This time dependence occurs in the regime corresponding to
Fig.\,2D. ({\bf A}) Surface plot of electrical transport for
$V_t=-151$ mV. Instability on the resonant peak is visible as
sharp current spikes.  The sweep direction is from $+$ to
$-\Delta_{LR}$, for fields stepped from $-$ to $+B_{ext}$. ({\bf
B}) Explicit time dependence of the resonant current exhibits
bistability ($V_t=-141$ mV, $B_{ext}=100$ mT). ({\bf C}) Lower
axis: dynamic nuclear polarization due to electron transport
through the device ($V_t=-141$ mV, $\Delta_{LR}=0$, $B_{ext}=200$
mT), after initialization to zero polarization by waiting for 5
minutes with no voltage applied. Top axis: in order to measure the
nuclear spin relaxation time, we wait for the current to saturate,
switch off the bias voltage for a time $t_{rel}$, and then
remeasure the leakage current. An exponential fit gives a time
constant of $80\pm40$ s (measurements of these long timescales
result in large error bars, $\pm 20$ fA due to $1/f$ noise). ({\bf
D}) The field dependence of the resonant current is hysteretic in
sweep direction ($V_t=-149$ mV). Each trace takes $\sim 7$
minutes.}
\end{center}
\end{figure}

\end{document}